\documentclass{revtex4}

\usepackage{graphicx}
\newtheorem{err}{Conjecture}

\begin{document}


\title{Convergence and Refinement of the Wang-Landau Algorithm}

\author{Hwee Kuan Lee}
\affiliation{Department of Physics, Tokyo Metropolitan University, Hachioji, Tokyo 192-0397, Japan}

\author{Yutaka Okabe}
\affiliation{Department of Physics, Tokyo Metropolitan University, Hachioji, Tokyo 192-0397, Japan}

\author{D. P. Landau}
\affiliation{Center for Simulational Physics, University of Georgia, Athens, GA 30602, USA}

\begin{abstract}
Recently, Wang and Landau proposed a new random walk algorithm that can be very
efficiently applied to many problems. Subsequently, there has been numerous studies
on the algorithm itself and many proposals for improvements were put forward.
However, fundamental questions such as what determines the rate of convergence has
not been answered. To understand the mechanism behind the Wang-Landau method, we did
an error analysis and found that a steady state is reached where the fluctuations in
the accumulated energy histogram saturate at values proportional to 
$[\log(f)]^{-1/2}$.
This value is closely related to the error corrections to the Wang-Landau method. We
also study the rate of convergence using different ``tuning" parameters in the
algorithm.
\end{abstract}

\maketitle


\section{Introduction}

Computational methods have been used extensively for solving complex problems in the
past decades. In particular, in statistical physics equilibrium quantities of a
system with many degrees of freedom are measured. The framework of statistical
physics is formalized such that all equilibrium quantities can be derived from the
partition function,
\begin{equation}
{\mathcal Z}(T) = \sum_{\{\sigma\}} \mbox{e}^{-E(\sigma)/k_{\mbox{\tiny B}} T}
\end{equation}
$\sigma$ is the state of the system, $E$ is the energy
corresponding to $\sigma$, $k_{\mbox{\tiny{B}}}$ is the Boltzmann
constant and $T$ is the temperature. The summation is over all
possible states and the number of possible states is a colossal
number which cannot be enumerated. Nevertheless, computational
methods such as Monte Carlo techniques~\cite{landau} are used to
sample the partition function; in particular, Metropolis
importance sampling~\cite{metropolis} has achieved considerable
success. However, new techniques are emerging and are replacing
the Metropolis importance sampling especially near phase
transition boundaries where the Metropolis importance sampling
becomes inefficient. A class of new techniques, called the
generalized ensemble methods, such as the multicanonical
method~\cite{berg,lee}, the broad histogram method~\cite{oliveira}
and the flat histogram method~\cite{jswang,yamaguchi2}, were
developed based on re-writing the partition function as a sum over
energies
\begin{equation}
{\mathcal Z}(T) = \sum_{\{\sigma\}} \mbox{e}^{-E(\sigma)/k_{\mbox{\tiny B}} T}
            = \sum_{E} g(E) \left[\mbox{e}^{-E(\sigma)/k_{\mbox{\tiny B}} T}\right]
\label{eq:z}
\end{equation}
where the partition function is reduced from a sum over all states
to a sum over $\sim$$N$ energy levels. The partition function
would be tractable if the energy density of states $g(E)$ could be
calculated.

Recently, a systematic, iterative, random walk
method~\cite{wang1,wang2,landau1} has been proposed as one of the
generalized ensemble methods. Now generally known as the
Wang-Landau algorithm, it has received much attention in
literature and has been applied to a wide range of
problems~\cite{yamaguchi1,jain,rathore,yan}. There have also been
numerous proposed improvements and studies of the efficiency of
this
method~\cite{schulz,shell,yamaguchi,schulz1,zhou,yan1,dayal,trebst,virnau};
however, there are still many unanswered questions, e.g. what
determines the rate of convergence to the true density of states
and is there any universality behavior related to this algorithm?
In this paper, we attempt to quantify the mechanism behind the
Wang-Landau method and study the effects of using different
``tuning" parameters.

The Wang-Landau algorithm~\cite{wang1,wang2,landau1} is an
iterative process in which the density of states $g(E)$ is
modified by a factor $f_k>1$, and the refinement of the density of
states is assured with monotonically decreasing modification
factors, e.g. $f_{k+1} = \sqrt{f_k}$. For each time the energy
level $E$ is visited, $g(E)$ is multiplied by $f_k > 1$, and a
histogram of energy is accumulated concurrently. It was proposed
that the modification factor be decreased when the accumulated
histogram satisfies a certain flatness condition which we call the
stopping condition. In this paper we study the effects of
different modification factors and stopping conditions, and derive
an expression for the error term in the Wang-Landau method based
on generalizations of the modification factors and stopping
conditions.
We shall consider arbitrary sequences of modification factors,
$f_1 > \cdots f_k > f_{k+1} \cdots > 1$ and arbitrary sequences of
corresponding stopping conditions $\lambda_1, \lambda_2, \cdots$.
The stopping conditions $\lambda_1, \lambda_2, \cdots$ may or may
not be the histogram flatness condition; other stopping conditions
could be used, for example, stopping after some predetermined
maximum number of Monte Carlo steps, or stopping after some number
of times the random walker reaches the minimum energy state. This
generalization is needed for theoretical error analysis in the
next section.

\section{Theoretical Error Analysis}
In the Wang-Landau algorithm, for each visit to an energy level
$E$, the density of states at that energy level increases by a
factor $f_k > 1$, and the corresponding histogram increases by
one. Assume that initially the unknown density of states were set
to $1$, i.e. $g_0(E)=1$ $\forall$ $E$. The density of states at
the end of the $n$th iteration is given by
\begin{equation}
\log g_n(E) = \sum_{k=1}^n H_k(E) \log(f_k)
\label{eq:geprimi}
\end{equation}
where $H_k(E)$ is the accumulated histogram and $f_k$ is the
modification factor for the $k$th iteration. At this point, it is
important to realize that the relative values of $g(E)$ are
sufficient for calculating thermodynamics quantities. Hence, a
constant factor can be extracted from $g_n(E)$, without losing any
information, by a change of variable on the histograms,
\begin{eqnarray}
H_k (E) \rightarrow H_k (E) - \min_E \{ H_k (E) \} = \tilde H_k (E) &
\mbox{ for } & k = 1, 2,\cdots, n
\label{eq:hefact}
\end{eqnarray}
where $\min_E \{H_k(E)\}$ is the minimum value of the accumulated histogram for the
$k$th iteration. Then Eq. (\ref{eq:geprimi}) becomes
\begin{equation}
\log g_n(E) = \sum_{k=1}^n \tilde H_k(E) \log(f_k) +
\mbox{constant of energy}. \label{eq:geconst}
\end{equation}
The second term in Eq. (\ref{eq:geconst}) is independent of
energy, and from here onwards we shall refer to $\tilde g_n(E)$ as
the density of states without the second term in Eq.
(\ref{eq:geconst}), i.e.
\begin{equation}
\log \tilde g_n(E) = \sum_{k=1}^n \tilde H_k(E) \log(f_k) .
\label{eq:gewl}
\end{equation}
To lay the foundation for deriving an expression for the error of
the Wang-Landau method, we use the conjecture that the method
converges to the true density of states with proper choice of
parameters.
\begin{err}
Let the Wang-Landau algorithm be carried out with a sequence of
modification factors, $ \cdots  f_k > f_{k+1} > \cdots
f_{\infty}=1$. There exists a sequence of stopping conditions
$\lambda_1, \lambda_2, \cdots \lambda_\infty$ such that,
\begin{equation}
\lim_{n \rightarrow \infty}
\tilde g_n (E) =
\tilde g_\infty (E) =
g^*(E) \times \mbox{constant}
\end{equation}
\label{con:only} where $\tilde g_n(E)$ is the density of states
calculated up to the $n$th iteration and $g^*(E)$ is the true
density of states.
\end{err}
This conjecture does not give any error bounds on the density of
states; it only says that, in the limit of an infinite number of
iterations the Wang-Landau estimate converges to the true density
of states. In addition, no constraint is imposed on the stopping
conditions in the conjecture. The error term up to the $n$th
iteration can be defined as
\begin{equation}
\sum_E \left[ \log \tilde g_{\infty}(E) - \log \tilde g_n(E) \right] =
\sum_E  \sum_{k=n+1}^\infty \tilde H_k(E) \log(f_k)
\label{eq:geerr}
\end{equation}
An intuitive view of Eq. (\ref{eq:geerr}) is that, if an infinite
number of iterations were performed, the exact answer would be
obtained.  When $n$ iterations were done instead, the error of
$\tilde g_n(E)$ will be the sum of all the rest of the iterations
that were not carried out explicitly. Define
\begin{equation}
\Delta H_k = \sum_E \tilde H_k(E)
\label{eq:dhk}
\end{equation}
where $\Delta H_k$ is a measure of fluctuations in $\tilde H_k(E)$. By the assumption
of convergence series (implied by Conjecture \ref{con:only}), the order of
summations in Eq. (\ref{eq:geerr}) can be rearranged. Then, Eq. (\ref{eq:geerr})
becomes
\begin{equation}
\eta_n =
\sum_E  \left[ \log \tilde g_{\infty}(E) - \log \tilde g_n(E) \right] =
\sum_{k=n+1}^\infty \Delta H_k \log(f_k)
\label{eq:geerrfinal}
\end{equation}
Eq. (\ref{eq:geerrfinal}) shows that, assuming appropriate
stopping conditions, $\eta_n$ depends only on the fluctuation in
the histogram and the sequence of modification factors $f_k$. When
the values of $f_k$ are predetermined (e.g. $f_{k+1} =
\sqrt{f_k}$), $\Delta H_k$ becomes the only determining factor for
$\eta_n$.

\section{Results}

We investigate the Monte Carlo time dependence of $\Delta H_k$ for
each iteration with the Wang-Landau method, where the subscript
$k$ denotes the $k$th iteration. Simulations were performed on the
ferromagnetic Ising model and on the fully frustrated Ising model
with various system sizes. The Hamiltonian is
\begin{equation}
{\mathcal H} = - \sum_{\langle ij \rangle} J_{ij} \sigma_i \sigma_j
\end{equation}
where the sum is over nearest neighbors on a two dimensional
square grid and $\sigma_i$ takes the values $\pm1$. $J_{ij}=1$ for
the ferromagnetic Ising model, and for the fully frustrated Ising
model, $J_{ij}$ takes the value $-1$ for every alternate
horizontal nearest neighbors bonds and $+1$ otherwise.

\begin{figure}
\begin{picture}(400,300)(0,0)
\put(25,0){\scalebox{.5}{\includegraphics{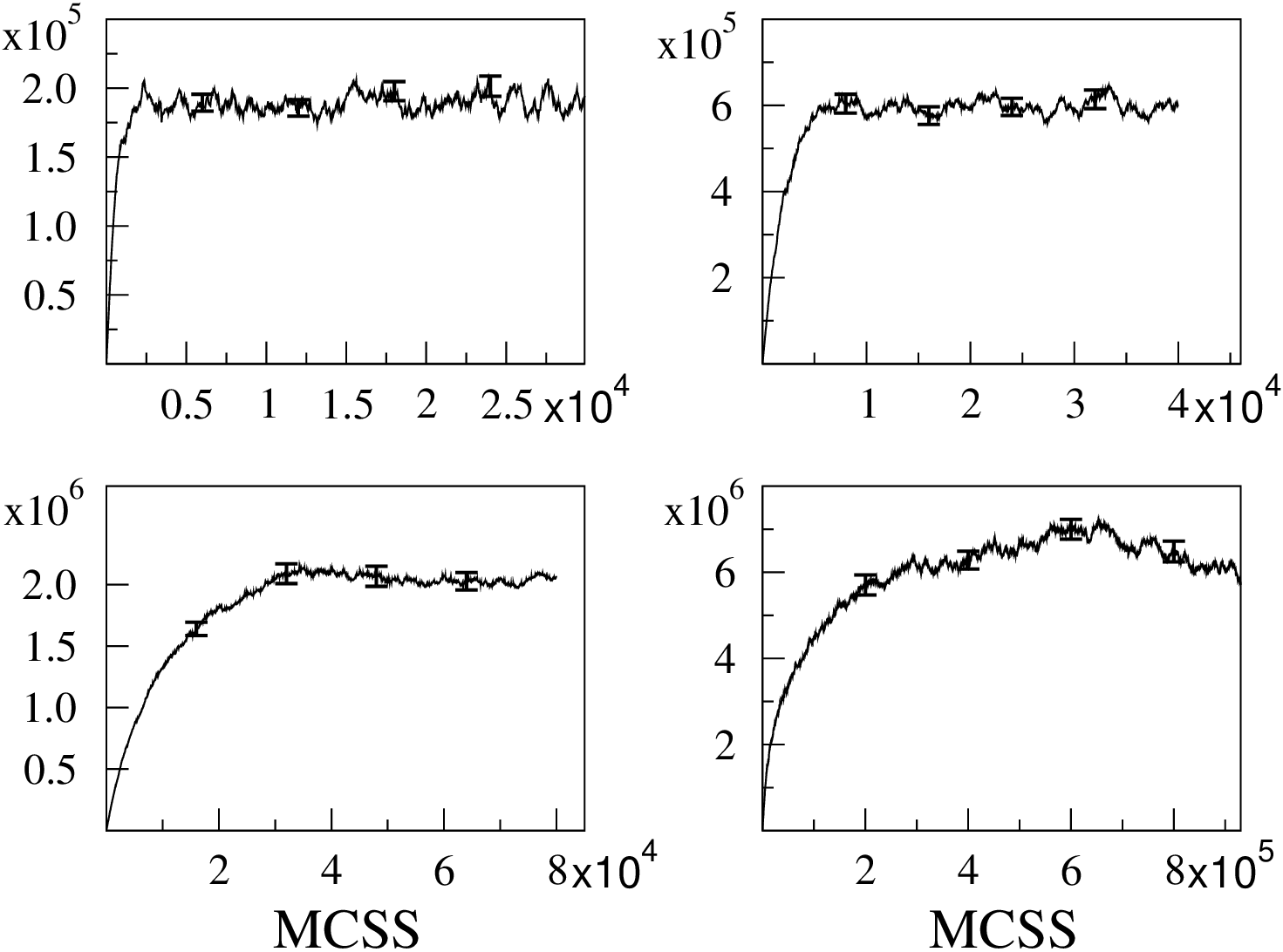}}}
\put(-5,220){\scalebox{1.4}{$\Delta H_k$}}
\put(-5,100){\scalebox{1.4}{$\Delta H_k$}}
\put(100,190){$\log(f_4)=10^{-2}$}
\put(270,190){$\log(f_8)=10^{-3}$}
\put(270,60){$\log(f_{16})=10^{-5}$}
\put(100,60){$\log(f_{12})=10^{-4}$}
\end{picture}
\caption{$\Delta H_k$ versus Monte Carlo steps per site for $16 \times 16$
ferromagnetic Ising model for various $\log(f)$ values. From left to right,
top to bottom, $\log(f)$
values are $10^{-2}$, $10^{-3}$, $10^{-4}$ and $10^{-5}$.}
\label{fig:satmcs}
\end{figure}
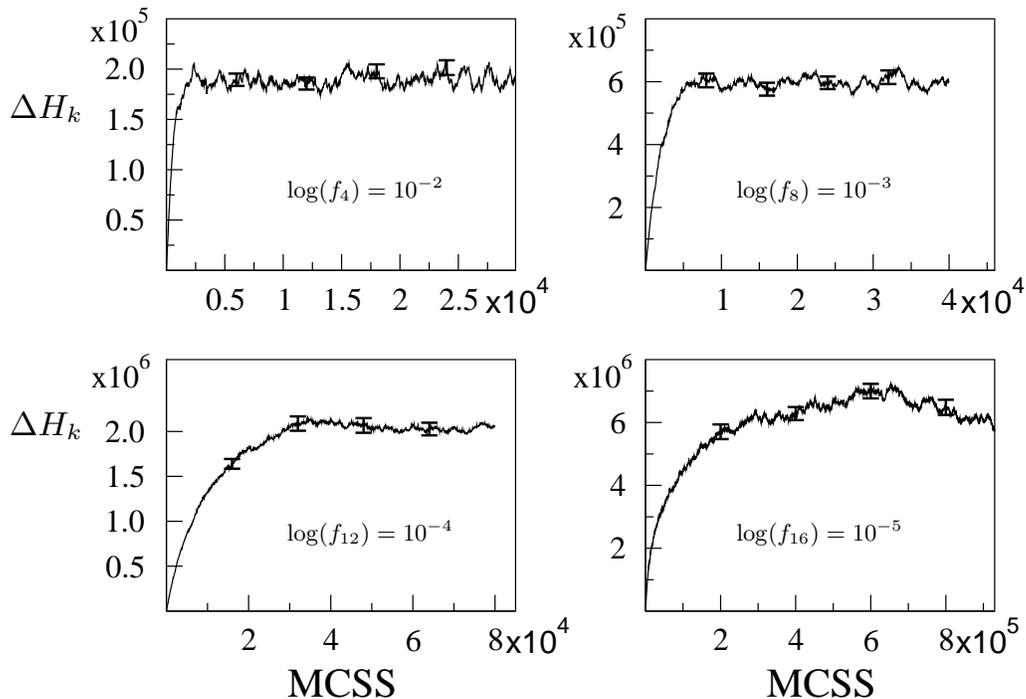

Fig. \ref{fig:satmcs} shows the time dependence of $\Delta H_k$ for several values of
$\log (f)$; $\log(f)=10^{-2}, 10^{-3}, 10^{-4}$ and $10^{-5}$ from left to right, top
to bottom respectively. We used the sequence of correction factors
$\log(f_{k+1})= \log(f_k)/ 1.78$ with $\log(f_1)=0.1$, this sequence is chosen so
that $\log(f_{k+4})=\log(f_{k})/10$. These graphs were generated by performing
the Wang-Landau algorithm on a $16 \times 16$ ferromagnetic Ising model with
numerical values averaged over 128 independent simulations. The Monte Carlo steps per
spin, the horizontal axis of Fig. \ref{fig:satmcs}, are measured from the time when
we decrease $\log(f)$. $\Delta H_k$ increases initially and eventually saturates,
and for smaller $\log(f)$ values, saturation values are greater and number of Monte
Carlo steps required to reach saturation are larger. Because the error term given by
Eq. (\ref{eq:geerrfinal}) depends only on $\Delta H_k$, any computational effort
after $\Delta H_k$ become saturated does not improve the accuracy of the final
density of states $g_n(E)$. On the other hand, stopping the random walk before
$\Delta H_k$ becomes saturated would make the simulation less efficient because
insufficient statistics are accumulated in the $k$th iteration and much more
statistics would have to be accumulated with smaller $\log(f)$ values for subsequent
iterations. An optimal algorithm is to stop the simulation as soon as $\Delta H_k$
becomes saturated. The Wang-Landau algorithm in the original paper~\cite{wang1}
suggested using the histogram flatness condition as a stopping condition, but this
does not guarantee optimal efficiency.

\begin{figure}
\begin{picture}(350,250)(0,0)
\scalebox{.4}{\includegraphics{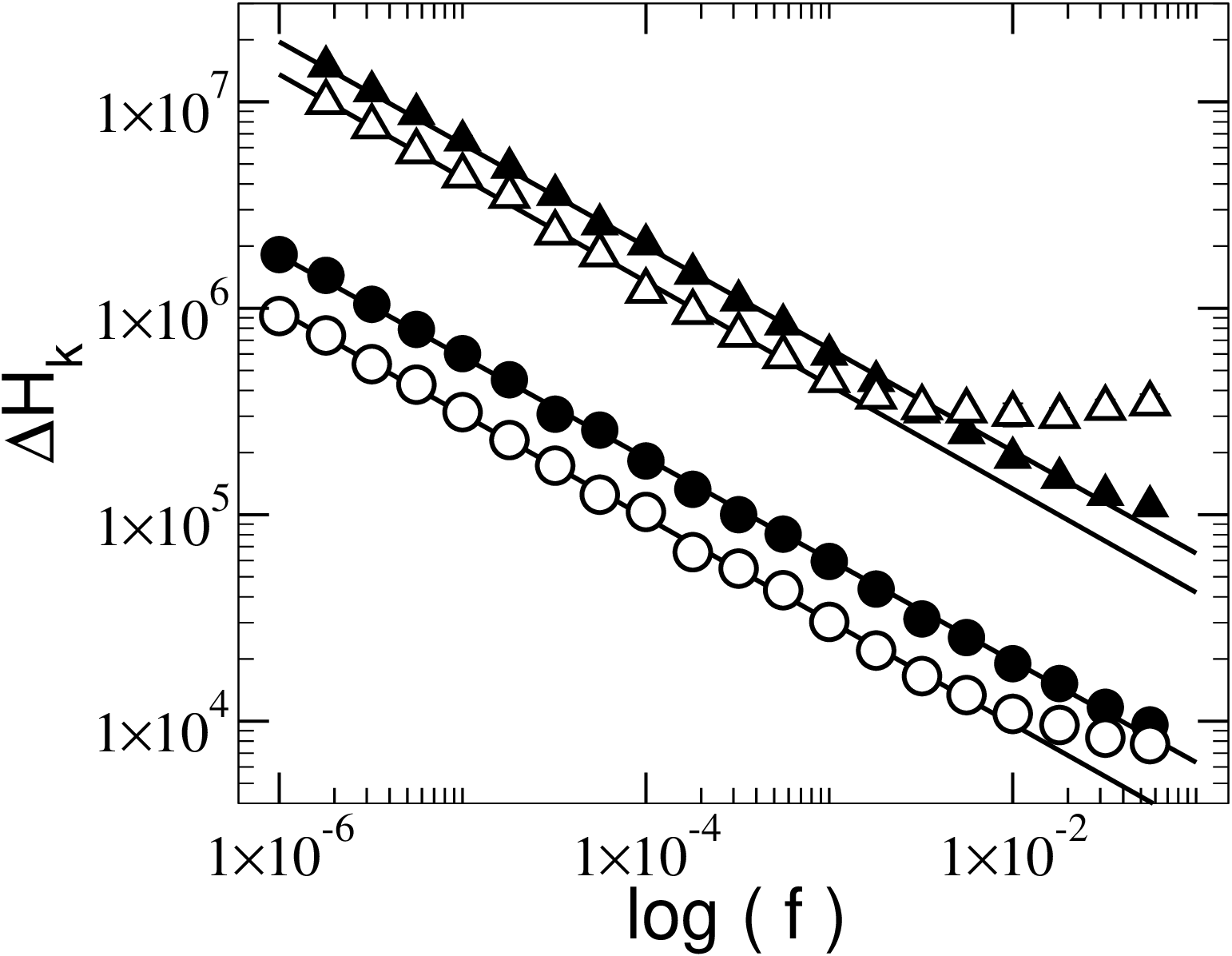}}
\end{picture}
\caption{Plots of saturation $\Delta H_k$ versus $\log(f)$ for $8 \times 8$
ferromagnetic Ising model (filled circles), $8 \times 8$ fully frustrated Ising model
(empty circles), $16 \times 16$ ferromagnetic Ising model (filled triangles) and
$16 \times 16$ fully frustrated Ising model (empty triangles). 128 independent
simulations were performed for each data point and error bars were smaller than the
size of the symbols.}
\label{fig:dH}
\end{figure}

It is difficult to predict the saturation value of $\Delta H_k$
for $k=1,\cdots$. As shown in Fig. \ref{fig:satmcs}, we performed
a set of simulations with more Monte Carlo steps than required for
$\Delta H_k$ to reach saturation. In this way, we could measure
the saturation values accurately. Fig. \ref{fig:dH} shows a plot
of saturation values versus $\log(f)$ for ferromagnetic Ising
model (FMIM) and fully frustrated Ising model (FFIM). In double
log scale, the data points fall on a straight line with the values
of the slopes equal to $-0.491 \pm 0.004$ for $8 \times 8$ FMIM,
$-0.501 \pm  0.004$ for $8 \times 8$ FFIM, $-0.496 \pm 0.006$ for
$16 \times 16$ FMIM and $ -0.502 \pm 0.008 $ for $16 \times 16$
FFIM. To within error bars, the slopes seem to have an universal
behavior
\begin{equation}
\max \{ \Delta H_k \} \propto  \log(f_k)^{-1/2} .
\end{equation}
as predicted by Zhou and Bhatt~\cite{zhou}. Our results suggest
that the values of the slope is generic to the Wang-Landau
algorithm and does not depend on system sizes and models.
Certainly many more simulations on different models are needed to
confirm the universality of the slope.
\section{Effects of Modification Factors}
We also looked at how the Wang-Landau algorithm performs with different sequences of
modification factors. In the extreme case, we studied the effects of taking the limit
of $f=1$ only after a few iterations. Assuming $n$ iterations were performed with
large modification factors, and on the $(n+1)$th iteration, the modification factor
is set to $1$. The background for implementation is as follows:
Eq. (\ref{eq:z}) uses the Boltzmann weights ($B(E,T) = 
\mbox{exp}(-E/k_{\mbox{\tiny B}} T)$) and the resulting energy distribution is,
\begin{equation}
P(E) = g^*(E) B(E,T) / {\mathcal Z}  = 
g^*(E) \mbox{exp}(-E/k_{\mbox{\tiny B}} T) / {\mathcal Z}
\end{equation}
where $g^*(E)$ is the true density of states. In general other weights can be used
in summing the partition function. If one chooses $B(E,T) = 1/ g_n(E)$, then
the probability distribution of $E$ for the $(n+1)$th iteration
$P_{n+1}(E)$ will be given by,
\begin{equation}
P_{n+1}(E)  = g^*(E) / g_{n}(E) {\mathcal Z}
\label{eq:pe}
\end{equation}
where $g_{n}(E)$ is the density of states
calculated by the $n$th iteration. $\mathcal Z$ is is an undetermined normalization
constant. The true density of states can then be estimated by the accumulated
histogram of the $(n+1)$th iteration.
\begin{equation}
g_{n+1}(E) =H_{n+1}(E) g_{n}(E) \times \mbox{constant}
\end{equation}
\begin{figure}
\begin{picture}(300,250)(0,0)
\scalebox{.4}{\includegraphics{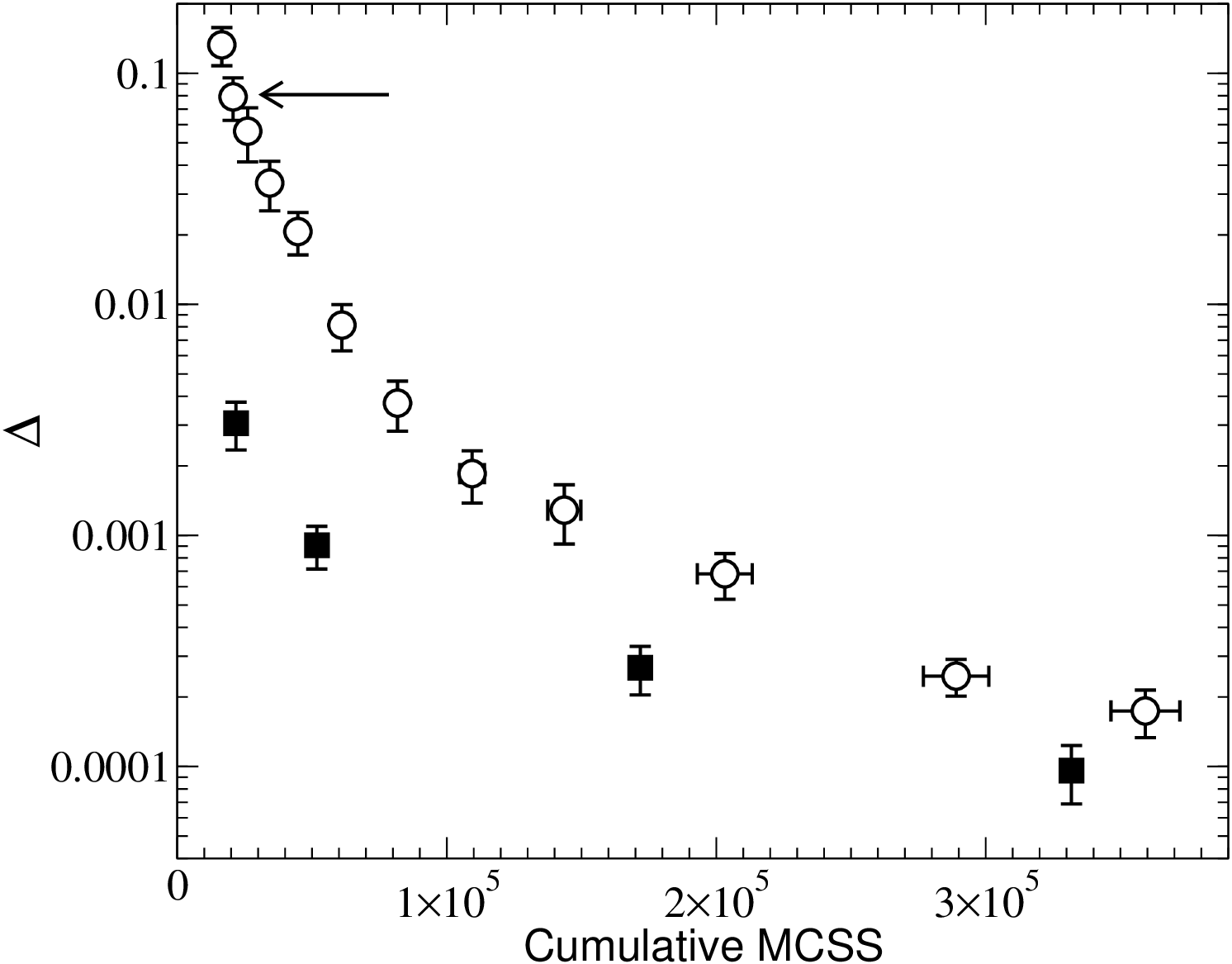}}
\end{picture}
\caption{Comparison of accuracy of the Wang-Landau method with two modification
sequences. Data points were generated with the sequence $f_{k+1}=\sqrt{f_k}$ (empty
circles) and with the sequence where the limiting value of $f=1$ is used after 
14 iterations (filled squares). Lattice size is $32 \times 32$ with energy range
$E/N_E \in [-1.55,-1.35]$.}
\label{fig:refined}
\end{figure}
This is analogous to the iteration process employed in Lee's
entropic sampling~\cite{lee}, but we used the fast diffusion of
the Wang-Landau algorithm in the early stage. Fig.
\ref{fig:refined} compares the accuracy of the Wang-Landau method
with two different modification sequences for the $32 \times 32$
ferromagnetic Ising model. The energy range was $E/N_E \in [
-1.55, -1.35]$ where $N_E=1024$ is the total number of lattice
sites. The vertical axis is the error of density of states defined
by,
\begin{equation}
\Delta = \frac{1}{m} \sum_E^m \left[ 1- \frac{g_{n+1}(E)}{g^*(E)} \right]^2
\end{equation}
where $g^*(E)$ is the exact density of states calculated from a MATHEMATICA program
provided by Beale~\cite{beale}. $g_{n+1}(E)$ is the calculated density of states and
$m$ is the total number of energy levels in the summation over this energy range. We
plot $\Delta$ for different sequences of modification factors. Empty circles were
generated with modification factors $f_{k+1} = \sqrt{f_k}$ (with $f_0=\mbox{exp}(1)$) 
and stopping when the condition 
$(H_{\mbox{\tiny max}}-H_{\mbox{\tiny min}})/
(H_{\mbox{\tiny max}}+H_{\mbox{\tiny min}}) \leq 0.1$ is satisfied. Where
$H_{\mbox{\tiny max}}$ and $H_{\mbox{\tiny min}}$ are the maximum and minimum 
histogram counts respectively.
Filled squares were
obtained with a sequence of modification factors where the limiting value of $f=1$
was used after 14 iterations. The arrow indicates the location which the
modification factor was set to $1$. We measure the errors (filled squares) at several
Monte Carlo steps per site after we set $f=1$. Error bars were obtained by averaging
over several independent simulations. Accuracy increases rapidly immediately after
setting $f=1$, but in the long run, the limiting case becomes only about twice as
accurate as the other sequence.

\section{Conclusion}

We derived an expression for the error term of the Wang-Landau algorithm. With
this, we showed that the fluctuation of the accumulated histogram
$\Delta H_k$ plays a central role in the accuracy of the Wang-Landau
method. We have also proposed that stopping each iteration as soon as
$\Delta H_k$ becomes saturated would be optimal.
The dependence of the saturation values on the
modification factor was also investigated and it was found that for the
ferromagnetic Ising model and fully frustrated Ising model,
%
$\max \{ \Delta H_k \} \propto  \log(f_k)^{-1/2}$.
%
With this equation, the saturation values of $\Delta H_k$ for
small modification factors can be predicted from values obtained
with larger modification factors. Perhaps, a more efficient
algorithm can be developed. We also studied the effects of using
different sequences of modification factors (refinement), in which
we presented the limiting case where the modification factor is
set to $1$ after 14 iterations. There are significant
improvements of efficiency for short simulations and improvements
become less for longer runs.

We wish to thank J. S. Wang for fruitful discussions. This work is
supported by the Japan Society for Promotion of Science, the
Laboratory Directed Research and Development Program of Oak Ridge
National Laboratory, DOE-OS through BES-DMSE and OASCR-MICS under
Contract No. DE-AC05-00OR22725 with UT-Battelle LLC, and by NSF
Grant No. DMR-0341874. The computation of this work has been done
using computer facilities of the Supercomputer Center, Institute
of Solid State Physic, University of Tokyo (Japan) and the
computer facilities of the Center for Computational Sciences, Oak
Ridge National Laboratory (U.S.A.).

\end{document}